\documentclass[pdflatex,sn-mathphys-num]{sn-jnl}% Math and Physical Sciences Numbered Reference Style
\usepackage{graphicx}%
\usepackage{multirow}%
\usepackage{amsmath,amssymb,amsfonts}%
\usepackage{amsthm}%
\usepackage{mathrsfs}%
\usepackage[title]{appendix}%
\usepackage{xcolor}%
\usepackage{textcomp}%
\usepackage{manyfoot}%
\usepackage{booktabs}%
\usepackage{listings}%
\usepackage[dvipsnames]{xcolor}
\theoremstyle{thmstyleone}%
\theoremstyle{thmstyletwo}%

\theoremstyle{thmstylethree}%

\begin{document}

\title[Article Title]{Astrophysical constraints on the cold equation of state of the strongly interacting matter}

\author[1,2]{\fnm{G\'abor} \sur{Kasza}}\email{kasza.gabor@wigner.hun-ren.hu}

\author[1,3]{\fnm{J\'anos} \sur{Tak\'atsy}}\email{janos.takatsy@uni-potsdam.de}
\equalcont{These authors contributed equally to this work.}

\author*[1]{\fnm{Gy\"orgy} \sur{Wolf}}\email{wolf.gyorgy@wigner.hun-ren.hu}
\equalcont{These authors contributed equally to this work.}

\affil[1]{\orgdiv{Theory Department}, \orgname{HUN-REN Wigner RCP}, \orgaddress{\street{POB 49}, \city{Budapest}, \postcode{1525}, \country{Hungary}}}

\affil[2]{\orgdiv{Institute of Technology}, \orgname{MATE KRC}, \orgaddress{\city{Gy{\"o}ngy{\"o}s}, \postcode{H-3200}, \country{Hungary}}}

\affil[3]{\orgdiv{Institut für Physik und Astronomie}, \orgname{Universität Potsdam}, \orgaddress{\street{Haus 28, Karl-Liebknecht-Str. 24-25}, \city{Potsdam}, \country{Germany}}}

\abstract{At present, the only experimental access to the properties of cold, dense strongly interacting matter is provided by astrophysical observations. Neutron stars are the only known systems in the Universe that reach densities several times higher than normal nuclear density at nearly zero temperature, making them unique laboratories for studying dense matter. Since most neutron star observables are sensitive to the equation of state (EOS), observational data place stringent constraints on the EOS of strongly interacting matter.

In this work, we investigate constraints arising from perturbative QCD calculations at asymptotically high densities ($\rho\approx 40 \rho_0$), the mass of the heaviest observed neutron star (a black widow pulsar), NICER mass–radius measurements, and the tidal deformability inferred from the binary neutron star merger GW170817. We parametrize the EOS and allow its parameters to vary freely, using observational data to constrain the admissible parameter space. We find that neutron star observations significantly restrict the EOS of dense strongly interacting matter. While NICER has already provided measurements for five pulsars, the associated uncertainties remain relatively large. Within our modeling framework, we find that the existence of very massive neutron stars and constraints on the tidal deformability provide the most restrictive constraints on the EOS.
%In contrast, the existence of very massive neutron stars and constraints on the tidal deformability emerge as particularly powerful probes of the EOS.
}




\maketitle

\section{Introduction}\label{sec1}

One of the central goals in nuclear physics is to explore the properties and phase diagram of strongly interacting matter. At present, this matter can be probed experimentally by heavy ion collisions only in limited regions of the phase diagram: at very low baryon chemical potential (e.g., RHIC at Brookhaven and the LHC at CERN) \citep{PHENIX,STAR,ALICE} and around normal nuclear density (ordinary nuclear physics). Reliable theoretical calculations exist in the low baryon chemical potential regime, lattice Quantum Chromodynamics (QCD) providing first-principles results \citep{Aoki2006,Borsanyi2013}. At extremely high baryon chemical potentials (corresponding to densities of roughly $\rho\approx 40\rho_0$, with $\rho_0$ the nuclear saturation density), perturbative QCD becomes applicable \citep{Gorda:2022jvk}. However, at intermediate baryon chemical potentials—where the critical end point is expected to reside—the properties of strongly interacting matter remain largely unknown, both experimentally and theoretically.
\begin{figure}[h]
\centering
\includegraphics[width=0.9\textwidth]{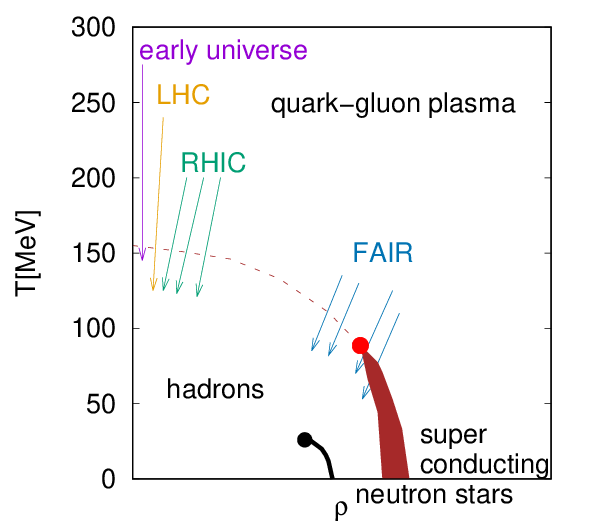}
\caption{The presumed structure of the phase diagram of strongly interacting matter.}\label{fig:phasediag}
\end{figure}

The forthcoming accelerator and detector facilities, such as NICA/MPD and FAIR/CBM, are expected to probe strongly interacting dense matter at temperatures of about 70–120 MeV \citep{HADES2019}. In contrast, at temperatures close to zero there is no terrestrial experiment capable of accessing the properties of dense matter. 

Currently, the only known system that reach densities several times higher than normal nuclear density at nearly zero temperature are neutron stars (NSs). Their observation therefore offers a unique opportunity to study cold, dense matter. Because most NS observables are sensitive to the equation of state (EOS) of dense matter, observational data place important constraints on the EOS. Several stringent constraints already exist, and more are expected in the near future \citep[e.g.][]{De:2018uhw, Tews:2018iwm,Rezzolla2017,Annala2017,Annala:2021gom,Pang:2021jta,Malik:2022zol,Altiparmak:2022bke,Marczenko:2022jhl,Koehn:2024set}. These arise from a variety of sources, including electromagnetic observations, gravitational-wave detections, and combined multi-messenger measurements.

Our strategy is the following: we parametrize the EOS. We let the parameters run freely and use the observations to provide constraints on the parameters of the EOS.

In Sec. 2, we describe our parametrization of the EOS. In Sec. 3, we summarize the experimental data used to constrain the EOS. Sec. 4 presents our results. Finally, in Sec. 5, we summarize our findings.

\section{Methods}

\subsection{Modelling the EOS}\label{sec:EOS}

%NS matter is highly isospin-asymmetric, which makes the symmetry energy a crucial ingredient of the EOS. Another key aspect is the role of strangeness at high densities. If strange baryons—such as the $\Lambda$ appear in NS matter, they modify the energy balance: introducing a $\Lambda$ increases the energy density but does not significantly raise the pressure, because the Fermi momentum of the nucleons is not increasing when some degrees of freedom are carried by strange baryons. As a result, the EOS becomes softer, which makes it challenging to support heavy NSs with masses $M_N > 2 M_odot$. Therefore, understanding the behavior and interactions of strange baryons is essential for accurately modeling NSs.

NSs just after their creation cool down temperatures below 1 MeV which can be considered to be zero in the strong interaction scale.
At low density, the cold strongly interacting matter is constrained by the nucleon-nucleon scattering data. %For strange baryon–nucleon interactions, one can rely on results from the HAL QCD collaboration \cite{HALQCD}cold  as well as $\Lambda(\Sigma)$–N femtoscopy measurements \cite{Femtoscopy}.
Around normal nuclear density, an enormous amount of information is available from conventional nuclear physics. This includes nuclear masses, isobaric analog states, hypernuclei, giant dipole and pygmy resonances, nuclear dipole polarizability, and neutron skin thickness. Together, these observables tightly constrain the equation of state (EOS) of strongly interacting matter up to saturation density $\rho_0$ and somewhat beyond.

At the opposite end of the baryon chemical potential axis, corresponding to densities of order $\rho \sim 40\rho_0$, T. Gorda et al. \cite{Gorda2018,Gorda2021,Gorda2023} computed the pressure, $p = 3.8\ \text{GeV/fm}^3$, at a chemical potential $\mu = 2.6\ \text{GeV}$ using a perturbative QCD (pQCD) calculation of order $\mathcal{O}(\alpha^3_s \ln\alpha_s)$ within the hard thermal loop approximation. Although such densities are not realized in nature, this result is important because the EOS of strongly interacting matter must connect to this regime causally, i.e., the speed of sound must not exceed the speed of light and the pressure must be a non-decreasing function of density.

Heavy-ion collision may provide information about the EOS at densities between these regimes, however, the temperature during these collisions is at the order of 100 MeV, and there are many competing effects, therefore, it is not straightforward how to gain information about the cold EOS from heavy-ion data.

A very exciting and promising possibility is to constrain the EOS of dense, strongly interacting matter by NS data. The aim of this paper is to investigate how the NS observations constrain the EOS of strongly interacting matter.

Below densities of approximately $0.5\rho_0$, the equation of state of nuclear matter is strongly affected by nuclear clustering effects. In this density regime, we therefore attach a dedicated crust equation of state to the hadronic EoS. Specifically, we employ the crust EoS derived from the unified model of Douchin and Haensel.~\cite{Douchin:2001sv}

In this work, for hadronic matter we employ the SFHo EOS based on the relativistic mean-field model of Steiner et al. \cite{Hempel2009,Steiner2013}, which includes nucleons and the $\omega, \rho$, and $\sigma$ mesons with quartic couplings. This EOS is relatively soft, for symmetric nuclear matter at saturation density with incompressibility $K=245$~MeV, slope of symmetry-energy $L=47.1$~MeV at saturation density, and the logarithmic derivative of the symmetry energy, and effective mass ratio $m^\star/m_n =0.76$ which are approximately the same for protons and neutrons. In addition to the saturation properties, it predicts binding energies and charge radii for $^{208}$Pb and $^{90}$Zr that are within 2\% of the experimental values. In this work, we neglect strangeness degrees of freedom, since strange baryons such as $\Lambda$ and $\Sigma$ appear to experience strong repulsive interactions at high densities \citep{Haidenbauer2017}. In a forthcoming study, however, we plan to include these degrees of freedom as well.

At higher densities, we employ the quark–meson model based on $U(3)\times U(3)$ chiral symmetry \citep{Parganlija:2012fy}. This model includes (pseudo)scalar and (axial-) vector meson nonets, constituent quarks, and Polyakov loops. Its parameters are fitted to vacuum hadron properties, such as masses and decay widths. The model reproduces lattice QCD results at zero chemical potential \citep{Kovacs:2016juc} and remains consistent with the lattice data up to baryon chemical potentials of $\mu \le 400$ MeV \citep{Kovacs:2017juc}. More details can be found in Refs.~\citep{Kovacs:2016juc,Takatsy2023}. Note that while in \citep{Takatsy2023} several different $m_\sigma$ values were considered, here in this study we only use $m_\sigma= 290$ MeV, as the best fit found in \citep{Kovacs:2016juc}. Since the vector coupling, $g_V$ is not set by the vacuum properties, and do not influence the thermodynamic properties of strongly interacting matter at $\mu=0$, we let it run free during the analyses.

To construct a unified EOS, these low- and high-density regimes must be connected. Since the two models involve different degrees of freedom, an effective method is required to describe the transition between them. Based on astrophysical observations, strong first-order phase transitions appear to be strongly constrained \cite{Christian:2020xwz,Gorda:2022lsk,Brandes2023}. Moreover, both models have limited ranges of validity, implying that at intermediate densities one must rely on an interpolation between the two. The simplest approach is to interpolate between the zero-temperature EOSs in an intermediate-density region (see, e.g., Refs.~\cite{Abgaryan2018,Baym2019,Masuda2012,Blaschke:2021poc}). As in earlier studies, we employ a polynomial interpolation; however, instead of using the pressure $p(\mu_B)$ as thermodynamic potential, we interpolate the energy density $\varepsilon(\rho_B)$. As shown in the following, this choice ensures the continuity of both the pressure and the speed of sound throughout the transition.

Assuming that the hadronic EOS $\varepsilon_\mathrm{H}(\rho_B)$, is valid up to a density $\rho_{BL}$, and that the quark EOS $\varepsilon_\mathrm{Q}(\rho_B)$ applies above $\rho_{BU}$, the interpolating EOS is written as
\begin{equation}
\varepsilon(\rho_B) = \sum_{k=0}^{N} C_k \rho_B^k , \quad \rho_{BL} < \rho_B < \rho_{BU},
\end{equation}
where the coefficients $C_k$ are chosen so that the energy density and several of its derivatives are continuous at $\rho_{BL} $ and at $\rho_{BU}$. In this work, we use a fifth-order polynomial which ensures continuity of the energy density and its first and second derivatives at the boundaries. The first derivative of $\varepsilon$ with respect to baryon density corresponds to baryon chemical potential, ensuring continuity of pressure, $p=\rho_B\mu -\varepsilon$. The continuity of the second derivative, in turn, ensures a continuous speed of sound: $c_s^2 = \frac{\partial p}{\partial \varepsilon}=\frac{\rho_B}{\mu} \frac{\partial^2\varepsilon}{\partial \rho_B^2}$. The coefficients $C_k$ can then be readily determined from these relations. From $\rho_{BL}$ and $\rho_{BU}$, we define the central density and width of the transition region as $\bar{\rho} \equiv (\rho_{BU}+\rho_{BL})/2$ and $\Gamma \equiv (\rho_{BU}-\rho_{BL})/2$, respectively.

We assume that the EOS derived from the quark–meson model remains valid up to the highest densities encountered in the cores of NSs. At even higher densities, no physical systems directly constrain the EOS; however, perturbative QCD calculations provide guidance in the asymptotically high-density limit. We therefore require that the EOS relevant for NS cores can be connected to the pQCD regime in a thermodynamically consistent way.

\begin{figure}[h]
\centering
\includegraphics[width=0.49\textwidth]{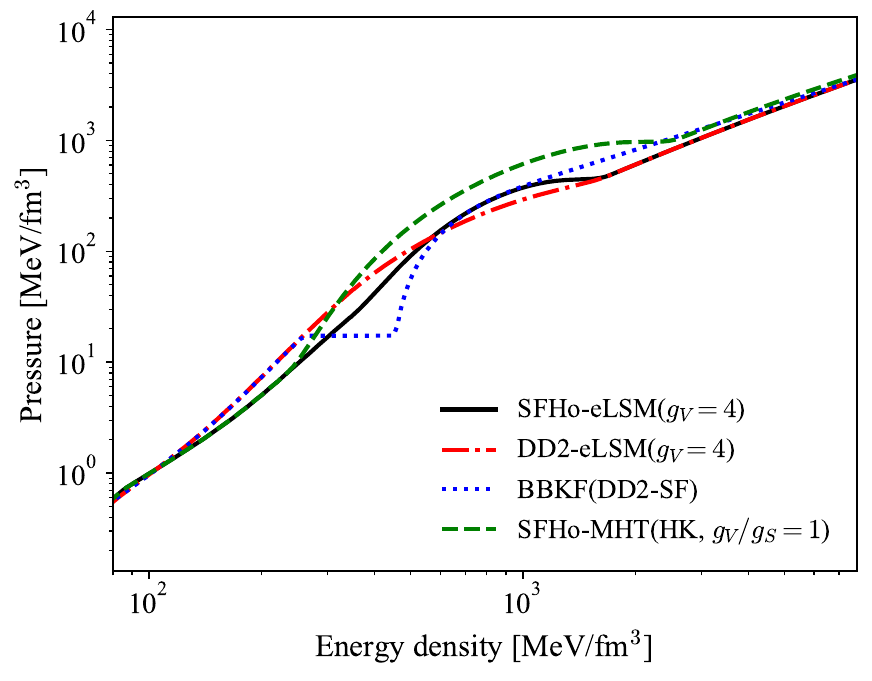}
\includegraphics[width=0.49\textwidth]{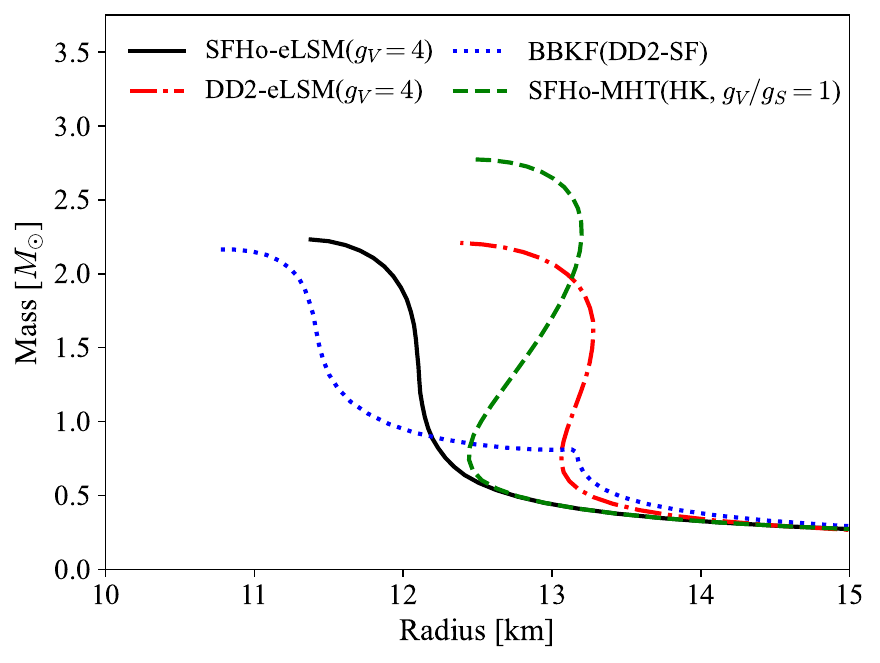}
\caption{Left panel: Equations of state considered in this work, including polynomially matched hadron–quark constructions based on the SFHo and DD2 hadronic EoSs, as well as an equation of state featuring a first-order phase transition. Right panel: Mass–radius relations of NSs calculated from the corresponding equations of state.
}\label{fig:EoS_MR_comparisons}
\end{figure}

In addition to our baseline construction, we compare our equation of state with several alternative scenarios to assess the robustness of our results. First, we repeat the polynomial matching of our quark-matter equation of state (eLSM) using the DD2 hadronic EOS~\cite{Hempel:2009mc,Typel:2009sy} instead of SFHo. Second, we consider an alternative description of quark matter based on the model of Masuda, Hatsuda, and Takatsuka (MHT)~\cite{Masuda:2012ed}, which is polynomially matched to the SFHo hadronic EOS. While Ref.~\cite{Masuda:2012ed} presents unified equations of state obtained by matching the quark-phase (NJL) EOS to several different hadronic models, in the present work we adopt only their quark-matter EOS and consistently match it to the SFHo hadronic EOS. This choice allows us to isolate and assess the impact of the quark-sector description by comparing unified equations of state that differ solely in the quark EOS, while keeping the hadronic sector fixed. Finally, we include the BBKF equation of state~\cite{Bastian:2020unt,Hempel:2009mc}, which features a first-order phase transition between the DD2 hadronic EOS and the string-flip (SF) quark-matter model. In all cases, the low-density crust is treated consistently with the procedure described above. The corresponding equations of state are compared in the left panel of Fig.~\ref{fig:EoS_MR_comparisons}, while their implications for NS structure are illustrated by the resulting mass–radius relations shown in the right panel of Fig.~\ref{fig:EoS_MR_comparisons}. In cases where polynomial matching is employed, we ensure that the resulting equations of state satisfy both stability and causality constraints. When our own quark-matter model is used for the quark phase, the matching parameters are fixed to $\Gamma = 2.8\rho_0$ and $\bar{\rho} = 5\rho_0$ in both constructions. In contrast, for the case based on the MHT model and the SFHo EOS, a different parametrization is required, with $\Gamma = 3.65\rho_0$ and $\bar{\rho} = 5.15\rho_0$.

\section{Theoretical and observational constraints}\label{sec:data}

\begin{itemize}
\item pQCD:

  Perturbative QCD (pQCD) becomes applicable at very high baryon densities because QCD itself becomes weakly coupled at large momentum scales, a property known as asymptotic freedom. At these densities bulk thermodynamics can be computed perturbatively. Accordingly, the EOS is required to approach the pQCD result while remaining causal, i.e. with $c_s < 1$. Following the pQCD calculation of Gorda et al. \cite{Gorda2018,Gorda2021,Gorda2023}, which was performed at next-to-next-to-next-to-leading order (N${}^3$LO) using hard-thermal-loop resummation, we impose convergence toward the pQCD reference point
  \begin{equation}
    \mu_\mathrm{QCD} = 2.6~\mathrm{GeV}, \qquad \rho_\mathrm{QCD} = 6.47~\mathrm{1/fm}^3, \qquad p_\mathrm{QCD} = 3823~\mathrm{MeV/fm}^3.
  \end{equation}
  \textcolor{red}{}

\item Minimal TOV mass constraint: $M_\mathrm{TOV}^\mathrm{min}=2.22M_\odot$:

% PSR J0740+6620, with a mass of $2.08\pm0.07$~$M_\odot$, and a $95.4\%$ lower bound of $1.95$~$M_\odot$ \citep{Fonseca2021}. Since that pulsar mass and radius was measured by the NICER detector, too, therefore that pulsar is considered together with the other NICER data. Here we we take into account: PSR J0348+0432 with a mass \textcolor{Blue}{$2.01 \pm 0.04 M_\odot$}\citep{Antoniadis2013}, and PSR J1614-2230 with a mass \textcolor{Blue}{$1.908 \pm 0.016 M_\odot$} \citep{Demorest2010}
  The masses of NSs in binary systems can be measured with remarkable precision, for example through the Shapiro time delay. Over the past decade, several highly massive NSs have been observed, providing robust constraints on the stiffness of the (EOS) \citep{Demorest2010,Antoniadis2013,Cromartie2019,Romani:2022jhd}. Among these, the most massive known object is PSR J0952–0607, a black widow pulsar, with a measured mass of $2.35\pm0.17$~$M_\odot$ \citep{Romani:2022jhd}. However, in this work we do not impose the mass of PSR J0952–0607 directly; instead, we adopt a more conservative constraint. Based on a combined analysis of the high-mass NS population, which includes PSR J0952–0607, Ref.~\citep{Romani:2022jhd} infers a lower bound on the maximum NS mass of $2.22,M_\odot$ at the $1\sigma$ confidence level. As such, this choice should not be interpreted as a strictly observational limit. There is a newer analysis \citep{Romani2025}, where the minimal TOV mass increased to $2.27 M_\odot$. Since this work is still in a preprint stage, we use the older value of $2.22 M_\odot$. 
  There is another analysis by Fan et al. \citep{Fan:2024Maxmass}, they obtained very similar value, $2.25_{-0.07}^{+0.08} M_\odot$. In this work, we adopt $2.22 M_\odot$ as a lower limit on the maximum NS mass and require all candidate EOSs to support NSs at least this massive.

\item
  NICER:
  
  Unlike NS masses, radius measurements are extremely challenging. One very promising technique for constraining NS radii—and, consequently, the dense-matter equation of state—exploits hot spots that form on the stellar surface due to the pulsar mechanism, accretion flows, or thermonuclear bursts on the NS surface. Pulse-profile modeling of these hot spots is employed by the Neutron Star Interior Composition Explorer (NICER), an X-ray telescope installed on the International Space Station in 2017, to infer NS properties, in particular the mass M and radius R.
  
\begin{figure}[h]
\centering
\includegraphics[width=0.9\textwidth]{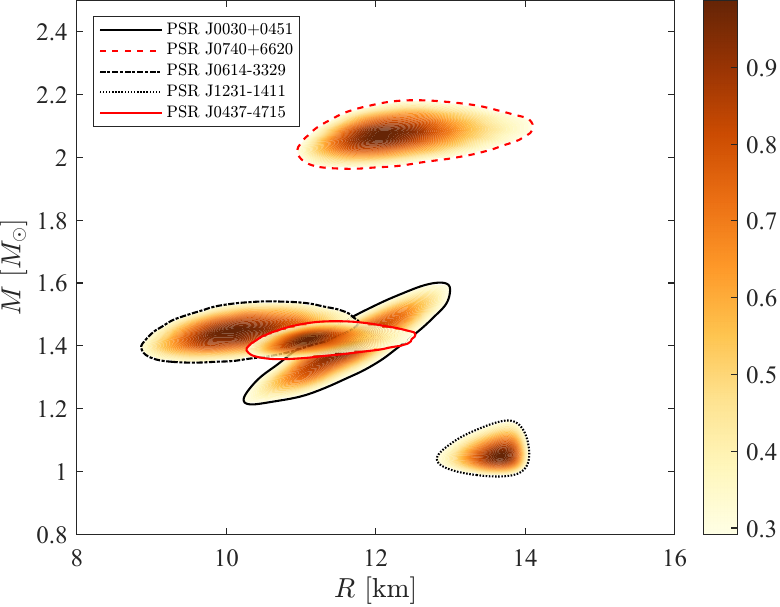}
\caption{The NICER results for the five pulsars with posterior distribution only within the contour 68\% \citep{Vinciguerra2024,Salmi2024b,Mauviard2025,Salmi2024,Choudhury2024}.
}\label{fig:NICER}
\end{figure}
To date, the most accurate X-ray measurements have achieved a precision of approximately $10\%$. Recent NICER analysis utilize the innovative approach of modeling the rotation-resolved X-ray spectra of NSs with surface hot spots. This method enables, for the first time with such a precision, a simultaneous measurement of the mass and radius of an individual NS. Using this technique, NICER has so far reported measurements for five NSs, which are briefly summarized below.

In a refined analyses  of Vinciguerra et al. \citep{Vinciguerra2024}, they obtained the mass and equatorial radius ($M$,$R$) values $1.4^{+0.13}_{-0.12}$~$M_\odot$ and $11.71^{+0.88}_{-0.83}$~km for PSR J0030+0451 with the so called ST-PDT model. Earlier analysises, \citep{Riley2019,Miller2019}, derived similar results. For the heavy pulsar, PSR J0740+6620 we use the results of $2.0073\pm0.069$~$M_\odot$ and $12.49^{+1.28}_{-0.88}$~km from \citep{Salmi2024b}. Another analysis reports similar radius  $12.76^{+1.49}_{-1.02}$~km \citep{Dittmann2024}. Three new pulsar data was published recently: PSR J0614-3329 with $M=1.44^{+0.06}_{-0.07}~M_\odot$ and $R_\mathrm{eq} = 10.29^{+1.01}_{-0.86}$~km \citep{Mauviard2025}, PSR J1231-1411 with $M=1.04^{+0.05}_{-0.03}$~$M_\odot$ and $R_\mathrm{eq} = 13.5^{+0.3}_{-0.5}$~km from Ref.~\citep{Salmi2024} (there is an alternative analysis with slightly different results \citep{Qi2025}), and PSR J0437-4715 with $M=1.418\pm 0.037~M_\odot$ and $R_\mathrm{eq} = 11.36^{+0.95}_{-0.63}$~km \citep{Choudhury2024} (here there is another analysis in a preprint, too \citep{Miller2025} with somewhat different results)
(Fig. \ref{fig:NICER}).
  
\item GW170817:

The first multimessenger observation of a binary NS merger, GW170817, provided important constraints on the NS equation of state. The initial analysis by the LIGO–Virgo Collaboration (LVC) inferred an upper bound of $\Lambda_{1.4}<800$ in the low-spin limit \citep{LIGOScientific2017}. Using generic EOS parameterizations, this was translated into an upper radius limit of R${}_{1.4}\le$13.6–13.7 km \citep{Annala2017,Most:2018hfd}. A subsequent LVC analysis combining tidal deformability and radius information, assuming a single EOS for both components, found $\Lambda_{1.4}=190_{-120}^{+390}$ and $R_{1.4}=10.8_{-1.7}^{+2.0}$ km \citep{LIGOScientific:2018cki}. 

An EOS-agnostic LVC analysis, exploring different waveform models under minimal assumptions, reported a conservative upper bound of $\tilde{\Lambda}<720$, where $\tilde{\Lambda}$ is the leading tidal contribution to the GW phase evolution, determined as a mass-weighted linear combination of the dimensionless tidal deformabilities of the merging NSs. \citep{LIGOScientific:2018hze}. This parameter is defined such that, for equal-mass inspiraling NSs with masses $M_1 = M_2$, their tidal deformability parameters $\Lambda_1$ and $\Lambda_2$ are equal to $\tilde{\Lambda}$~\citep{Wade:2014vqa}. With these definitions in mind, in this work, we adopt the conservative upper bound $\tilde{\Lambda}<720$\footnote{For a given EOS we calculate all the possible $\tilde{\Lambda}$ values for primary masses $1.362 M_{\odot}<M_1<1.6 M_{\odot}$, and keep the EOS if $\tilde{\Lambda}<720$ for any configuration. The lower limit of $M_1$ was determined based on the chirp mass, whose value ($\mathcal{M} = 1.186 M_{\odot}$) was given in Ref.~\cite{LIGOScientific:2018hze}.} from the EOS-agnostic LVC analysis of GW170817, whose minimal prior assumptions make it well suited for this purpose.

We also examine the impact of imposing a more stringent tidal-deformability constraint,
\begin{equation}
  70 < \Lambda_{1.4} < 580, 
\end{equation}
together with a radius constraint,
\begin{equation}
  9.1\textnormal{ km} < R_{1.4} < 12.8\textnormal{ km}.
\end{equation}
These bounds correspond to $\Lambda_{1.4} = 190^{+390}_{-120}$ and $R_{1.4} = 10.8^{+2.0}_{-1.7}\,\mathrm{km}$ at the 90\% confidence level, as reported in Ref.~\citep{LIGOScientific:2018cki}.

%\item
%  \textcolor{Magenta}{hypermassive NS} GW170817: 
%  no prompt black hole formation,  (Rezzola): \textcolor{Blue}{$2.01\ge M_{TOV}/M_\odot \ge 2.16$} %$M_{thres}< 2.53 M_\odot$ 
\item HESS:

A recent analyses reported a very light central compact object in the supernova remnant HESS J1731–347, interpreted as the lightest NS observed to date, with an inferred mass of $0.77^{+0.20}_{-0.17}$~$M_\odot$ and radius $10.4^{+0.86}_{-0.78}$~km. However, the interpretation of this object as a NS and the robustness of the inferred properties are generally not accepted \citep{Alford:2023waw}, we still examine as a hypothetical scenario, whether such a NS is allowed by our model.
  
\item Mass gap NS:

  An object in the theorized lower mass gap between observed NSs and black holes was detected in the GW event GW190814, with an inferred mass of $M=2.59^{+0.08}_{-0.09}$~$M_\odot$ at the 90\% credible level \citep{LIGOScientific:2020zkf}. In our analysis, we also explore the implications of interpreting this object as a NS. To this end, we apply a mass constraint using the error-function approach, adopting a mean value of $M=2.59$~$M_\odot$ and a standard deviation of $\Delta M=0.055$~$M_\odot$.
  
\end{itemize}

\subsection{Bayesian inference}

We use Bayesian inference to investigate the impact of the observational data described in Sec.~\ref{sec:data} on the manifold of possible equations of state (EOSs). The EOS constructed in Sec.~\ref{sec:EOS} is characterized by three parameters: the vector coupling $g_V$ between vector mesons and constituent quarks, the central density of the transition region $\overline{\rho} \equiv (\rho_{BU}+\rho_{BL})/2$, and its width $\Gamma\equiv (\rho_{BU}-\rho_{BL})/2$. We impose the following parameter ranges:
\begin{eqnarray} \label{eg:paramlimits}
    0&\le& g_V \le 10, \nonumber\\
   2 \rho_0 &\le& \overline{\rho} \le 5 \rho_0 \nonumber,\\
   \rho_0 &\le& \Gamma \le 4 \rho_0 \qquad \mathrm{with}\:  \mathrm{the}\: \mathrm{constraint:} \quad
   \rho_{BL} \equiv \overline{\rho} - \Gamma > \rho_0 .
\end{eqnarray}
For the Bayesian analysis, we assume a uniform prior probability distribution over the allowed parameter space. Using this prior, we construct approximately 10,000 EOSs that uniformly sample the parameter space.

In our analysis, observational constraints are incorporated in two different ways. In the first approach, we impose strict conditions which must be satisfied by an EOS in order to be retained. In the second approach, we impose probabilistic conditions, assigning likelihoods to EOSs based on how well they reproduce the data. As an example of a strict condition, we require all EOSs to support NSs with masses at least as large as the heaviest reliably observed NS, $M_\mathrm{TOV}^\mathrm{min}=2.22 M_\odot$.

To assign probabilities to EOSs, we use Bayes’ theorem. Let $\theta$ denote the parameter set defining an EOS, $p(\theta)$ the prior probability, and $p(\mathrm{data}|\theta)$ the likelihood of observing the data given $\theta$. The posterior probability is then
\begin{equation}
p(\theta|\mathrm{data}) =
\frac{p(\mathrm{data}|\theta)\:  p(\theta)}{p(\mathrm{data})},
\end{equation}
where $p(\mathrm{data})$ is a normalization constant. As stated above, we assume a uniform prior $p(\theta)$ over the allowed hypersurface. For statistically independent observations \citep{Takatsy2023}, the likelihood factorizes over the individual data sets. Let $\mathcal{D}_i = \{\mathrm{NICER},\, \mathrm{GW170817},\, \mathrm{HESS},\, \mathrm{GW190814}\}$ denote these independent observations. Hereafter, the GW190814 constraint will be referred to as the mass-gap likelihood, $p(\mathrm{Gap}|\theta)$. Then, the total likelihood can be written compactly as
\begin{equation}
p(\mathrm{data} | \theta) = \prod_i p(\mathcal{D}_i | \theta).
\end{equation}
It should be noted that, throughout the analysis, multiple configurations were considered, each employing only a subset of the available constraints, as discussed later.

The likelihoods for the individual measurements are calculated in the following manner. NS mass measurements are accounted for by:
\begin{equation}
    p(M_\mathrm{TOV}|\boldsymbol{\vartheta})= \frac{1}{2} \left[ 1 + \mathrm{erf}\left( \frac{M_\mathrm{TOV}(\boldsymbol{\vartheta}) - M_i}{\sqrt{2} \sigma_i} \right) \right] \: ,
\end{equation}
with $M_i$ the measured mass of the NS and $\sigma_i$ its uncertainty. Observations that provide PDFs for simultaneous mass-radius measurements, such as the NICER or HESS measurements, are used through the following integral formula:
\begin{equation}
    p(\mathrm{MR}|\boldsymbol{\vartheta}) \propto
    \int \mathrm{d}M \, p_\mathrm{N}(M,R=R(M,\boldsymbol{\vartheta})) \: .
\end{equation}
Finally, the tidal deformability measurement for GW170817 is incorporated as:
\begin{equation}
    p(\tilde{\Lambda}|\boldsymbol{\vartheta}) \propto \int\limits_{M_\mathrm{eq}}^{M_\mathrm{TOV}} \mathrm{d}M_1 \, p_\mathrm{GW}(\tilde{\Lambda}(M_1,\mathcal{M},\boldsymbol{\vartheta}),q(M_1,\mathcal{M})) \: ,
\end{equation}
where $M_1$ is the mass of the primary component of the binary NS, $M_\mathrm{eq}=1.362$~$M_\odot$ corresponds to a mass ratio of $q=1$, and $\mathcal{M}$ denotes the chirp mass of the binary \citep{LIGOScientific:2018hze}.

%As stated above, we assume a uniform prior $p(\theta)$ over the allowed hypersurface. For statistically independent observations \citep{Takatsy2023}, the likelihood factorizes as
%\begin{equation}
%p(\mathrm{data}|\theta)
%= p(\mathrm{NICER}|\theta) \:
%p(\Lambda_{1.4}|\theta)  \:
%p(M|\theta).
%\end{equation}

In addition to requiring that each EOS supports NSs with masses of at least $M_\mathrm{TOV}^\mathrm{min}=2.22~M_\odot$, we also impose as a strict condition that the EOS can be causally connected to the perturbative QCD (pQCD) point from the central density of the maximum-mass NS configuration. We implement this constraint according to Refs.~\citep{Komoltsev:2021jzg,Gorda:2022jvk}, and similarly to Ref.~\citep{Takatsy2023}. That is, we require causality and stability for the EOS, i.e.:
\begin{equation}
    n_{\text{NS}} \leq n_{\text{QCD}}\: , \quad
    p_\mathrm{NS} \leq p_\mathrm{QCD} \: ,
\end{equation}

\begin{equation}
    \frac{n_\mathrm{NS}}{\mu_\mathrm{NS}} \leq \frac{n_\mathrm{QCD}}{\mu_\mathrm{QCD}} \: ,
\end{equation}
while we also enforce the integral constraint
\begin{equation}
    \Delta p_\mathrm{min} \leq \Delta p \leq \Delta p_\mathrm{max} \: ,
\end{equation}
with $\Delta p \equiv p_\mathrm{QCD} - p_\mathrm{NS}$, and
\begin{equation}
    \Delta p_\mathrm{min} = \frac{\mu_\mathrm{QCD}^2-\mu_\mathrm{NS}^2}{2}\frac{n_\mathrm{NS}}{\mu_\mathrm{NS}} \:,
\end{equation}
\begin{equation}
    \Delta p_\mathrm{max} = \frac{\mu_\mathrm{QCD}^2-\mu_\mathrm{NS}^2}{2}\frac{n_\mathrm{QCD}}{\mu_\mathrm{QCD}} \: .
\end{equation}

All remaining observational constraints are treated probabilistically unless stated otherwise. In particular, NICER measurements provide probability distributions rather than sharp bounds. 
%Moreover, the mean mass and radius values inferred for different NSs by the NICER collaboration are not mutually consistent, even at the one $\sigma$ level, as illustrated in Fig.~\ref{fig:NICER}. 
Consequently, NICER data are incorporated exclusively through likelihood functions.

\section{Results}\label{sec:Results}

\begin{figure}[h]
\centering
\includegraphics[width=0.45\textwidth]{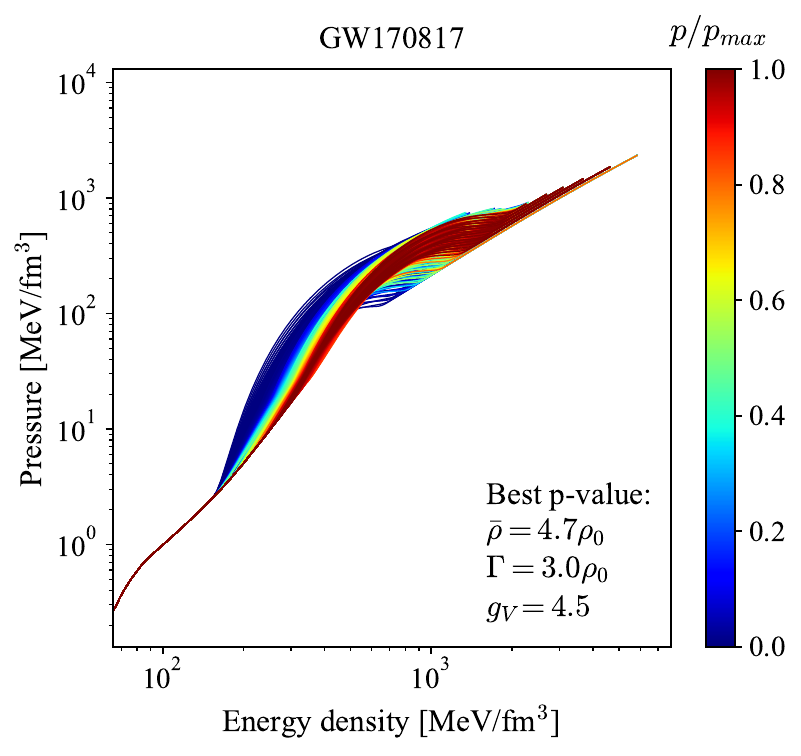}
\includegraphics[width=0.45\textwidth]{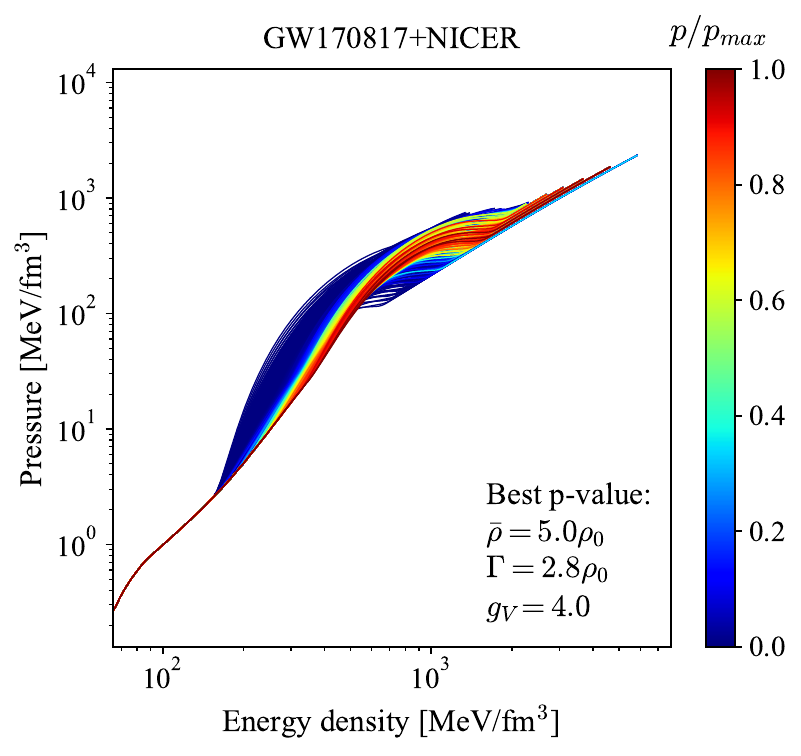} \\
\includegraphics[width=0.45\textwidth]{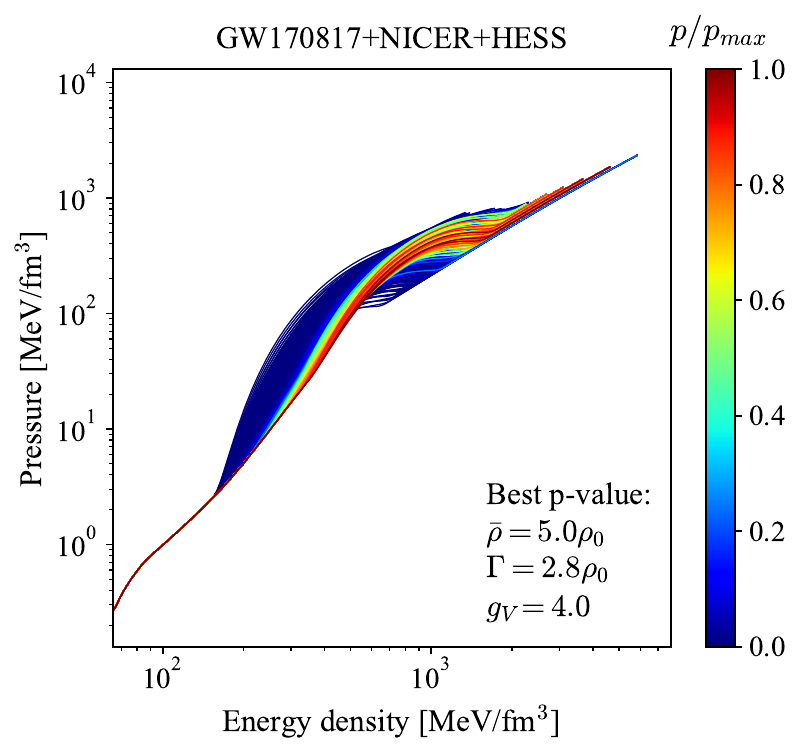}
\includegraphics[width=0.45\textwidth]{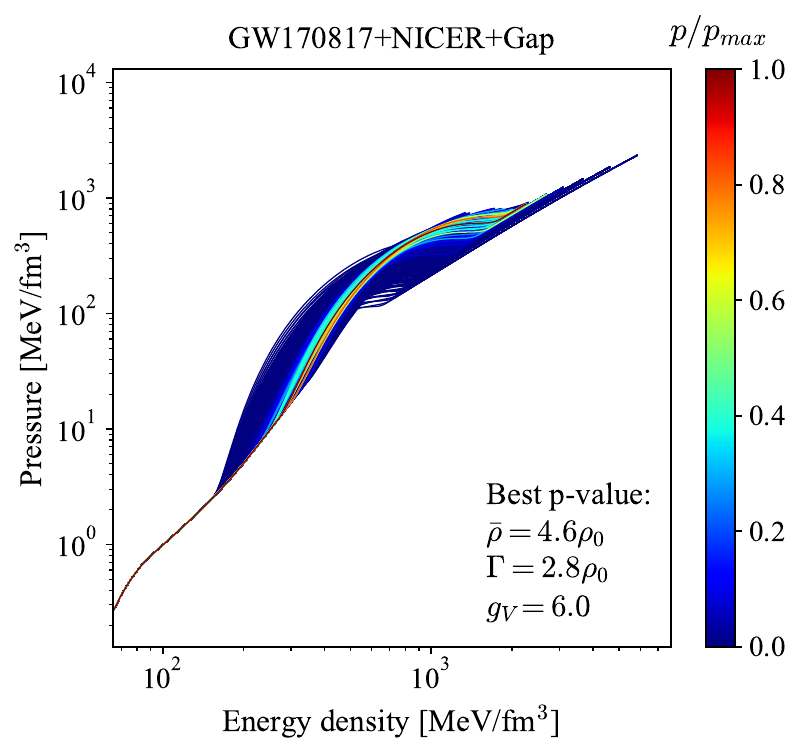}
\caption{The EOS curves coloured by their likelihood are shown when applied the minimal TOV mass and pQCD constraints with the GW condition (left upper figure); the GW and NICER conditions (right upper figure); the GW, NICER and the HESS conditions (left lower figure); GW, NICER, and the mass gap conditions (right lower figure). The most probable parameter set is indicated in each panel.}\label{fig:pQCD_BlWi-EOS}
\end{figure}

In Fig.~\ref{fig:pQCD_BlWi-EOS}, we show the probability distribution of EOSs obtained by imposing the strict conditions of the minimum-mass requirement and the pQCD constraint, while sequentially adding the remaining probabilistic constraints. At low energy densities $\varepsilon$, the EOS is well constrained by its hadronic component and appears as an approximately straight line. As the crossover region is reached, the slope increases significantly, the distribution broadens, and the EOSs become curved. At higher energy densities, the individual EOSs again approach nearly linear behavior, indicating that NS matter has transitioned into the quark-matter phase. Each curve is truncated at the point up to which it remains consistent with pQCD.

The allowed region for the EOS is rather narrow. Evidently, requiring the object in the mass gap to be interpreted as a NS would impose the strongest additional constraint on the EOS.

The analysis also yields the most probable parameter set. These values depend only weakly on the specific set of constraints applied. The most probable values are:
\begin{equation}
    4.6 \rho_0 \le \overline{\rho} \le 5 \rho_0, \quad 
    2.8 \rho_0 \le \Gamma \le 3.0 \rho_0 , \quad
    4 \le g_V \le 6 .
\end{equation}

To assess which observational constraint provides the strongest restrictions on the EOS, we determine how many EOSs from the original ensemble of the order of 10,000 satisfy each constraint when imposed as a strict condition.

\begin{table}[h]
\caption{The number of allowed EOSs are shown by applying more and more constraints.}\label{tab:NumEOS}%
\begin{tabular}{@{}|c|c|c|c|c|c|c|c|@{}}
\toprule
Initial set & Prior &$ +M_\mathrm{TOV}^\mathrm{min}$ & +pQCD  & +$\tilde{\Lambda}$ & +$\Lambda_{1.4}$ & +$R_{1.4}$ \\ %& $M_\mathrm{gap}$+pQCD+$\Lambda$\\
\midrule
9261  & 1993 & 593 & 438  & 133 & 104 & 104\\ % & 33 \\
\botrule
\end{tabular}
\end{table}

Table~\ref{tab:NumEOS} summarizes the number of EOSs that remain viable as increasingly stringent constraints are applied. We start from an initial set of 9,261 EOSs. Only about one fifth of these satisfy the basic stability and causality requirements; this subset is adopted as a hard prior, assigning unit probability to equations of state that satisfy these constraints and zero otherwise. Imposing the minimum-mass constraint,  $M_\mathrm{TOV}^\mathrm{min}=2.22M_\odot$, further reduces the number of allowed EOSs to approximately 600. Adding the pQCD constraint has a modest but non-negligible effect on this number.

The most significant reduction occurs when the tidal-deformability constraint \linebreak $\tilde{\Lambda}<720$ \citep{LIGOScientific:2018hze} is applied, which lowers the fraction of allowed EOSs to below  2\%. Imposing the more stringent tidal-deformability bound  $70 < \Lambda_{1.4} < 580$, associated with a NS of mass $1.4\,M_\odot$ \citep{LIGOScientific:2018cki}, further shrinks the viable EOS manifold. The corresponding radius constraint from the same analysis, 9.1 km $< R_{1.4} < $ 12.8 km, does not lead to any additional reduction once the tidal-deformability constraint is already enforced. This demonstrates that tidal deformability provides a particularly strong constraint on our model. We believe that this arises because the model we employ tends to favor larger radii, whereas the tidal deformability measurements generally point toward smaller radii. This is also illustrated by the observation mentioned above: imposing the constraint $9.1~\mathrm{km} < R_{1.4} < 12.8~\mathrm{km}$ does not further reduce the set of allowed equations of state when the constraint $70 < \Lambda_{1.4} < 580$ is already applied.

\begin{figure}[h]
\centering
\includegraphics[width=0.45\textwidth]{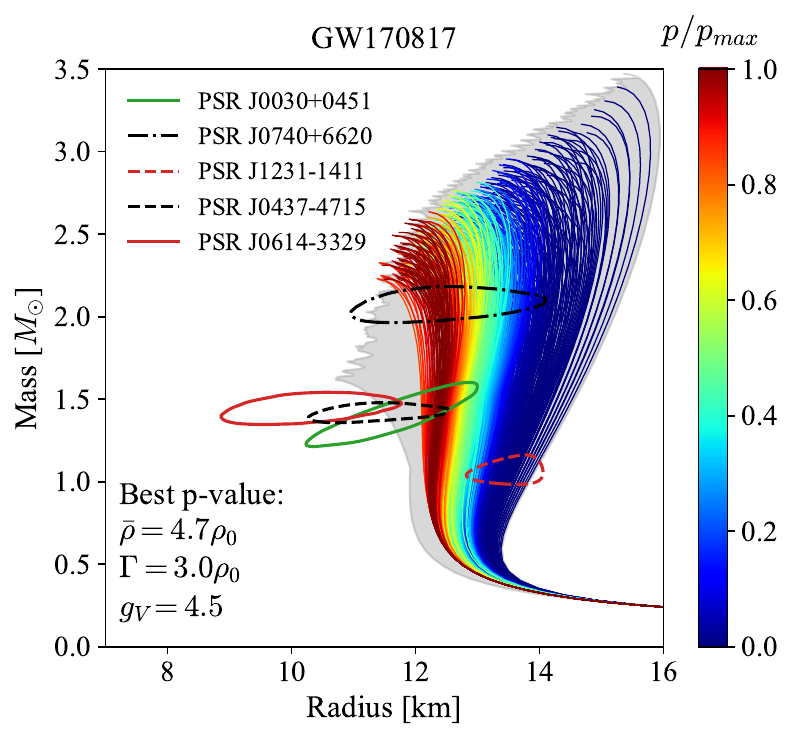}
\includegraphics[width=0.45\textwidth]{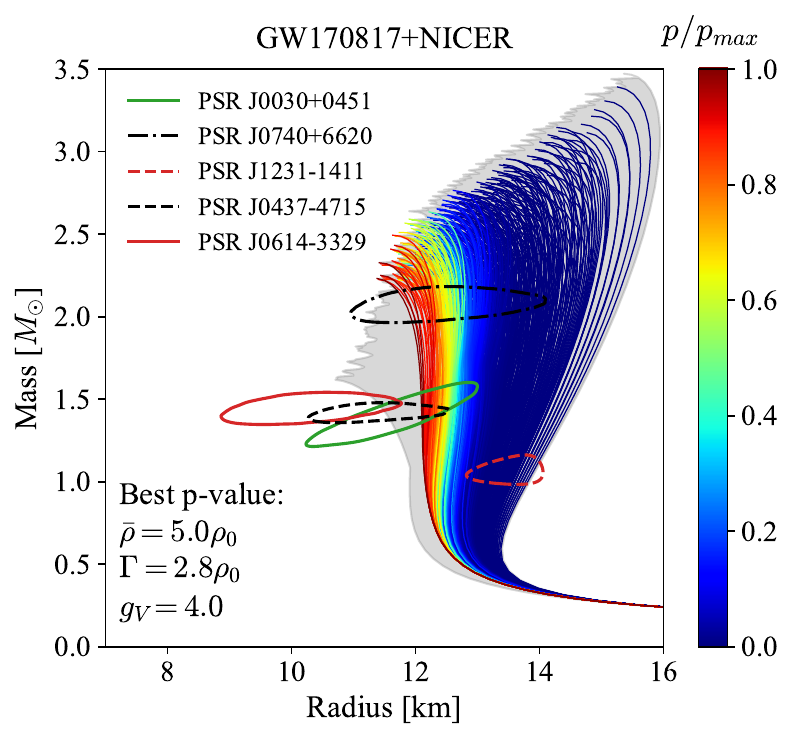} \\[3pt]
\includegraphics[width=0.45\textwidth]{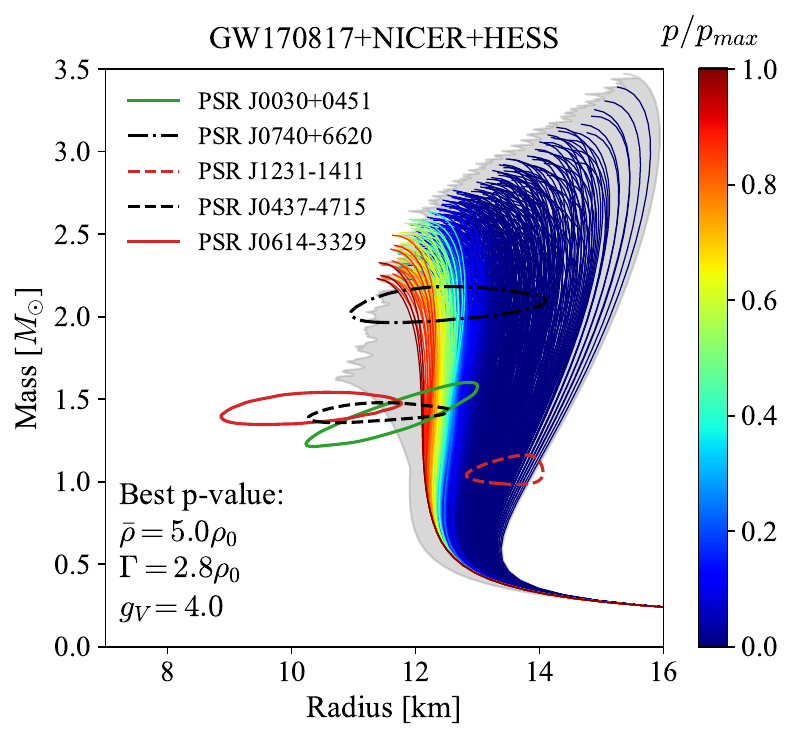}
\includegraphics[width=0.45\textwidth]{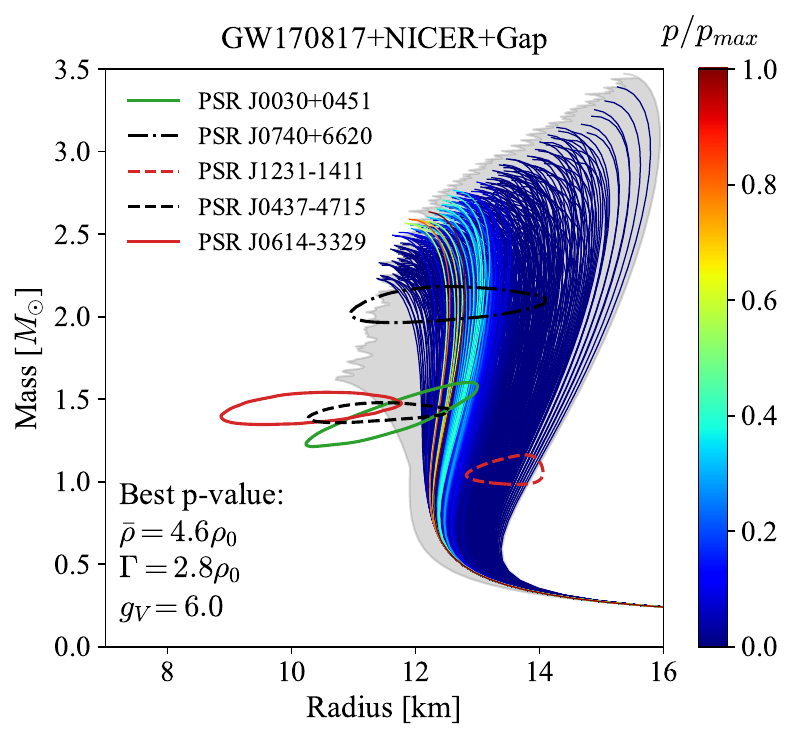}
\caption{The individual $M(R)$ curves coloured by their likelihood are shown when applied the minimal TOV mass and pQCD constraints with the GW condition (left upper figure); the GW and NICER conditions (right upper figure); the GW, NICER and the Hess conditions (left lower figure); GW, NICER, and the mass gap conditions (right lower figure). The grey shaded background indicates the region spanned by the prior, defined by the envelope of the corresponding family of curves. \label{fig:pQCD_BlWi-MR}}
\end{figure}

In Fig.~\ref{fig:pQCD_BlWi-MR}, we show the sample of the allowed $M(R)$ curves colored by their likelihood, with the 68\% confidence contours from NICER observations overlaid for comparison. The grey shaded background denotes the region covered by the prior, as defined by the envelope of the curve ensemble. The panels correspond to the same conditions as in Fig. \ref{fig:pQCD_BlWi-EOS}.

\begin{figure}[h]
\centering
\includegraphics[width=0.49\textwidth]{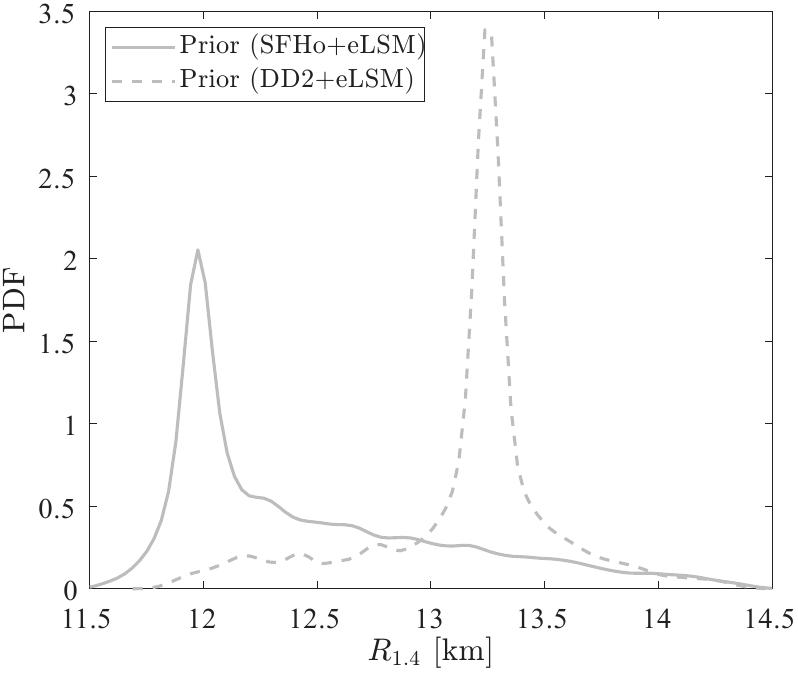}
\hfill
\includegraphics[width=0.49\textwidth]{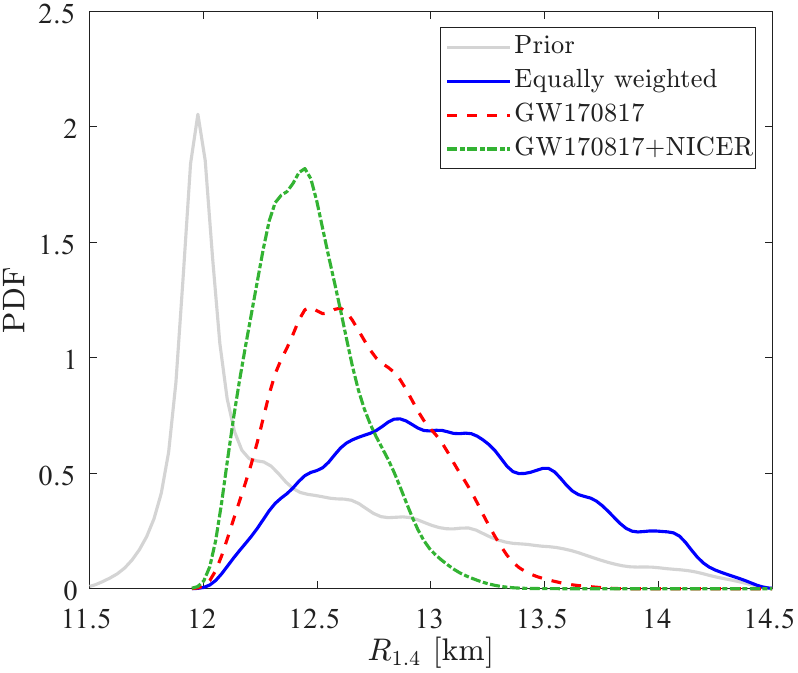}
\caption{\textit{Left panel}: Prior-induced distributions of the radius $R_{1.4}$ of a $1.4,M_\odot$ NS, obtained from the hard prior on the equation of state, shown for two different choices of the hadronic phase: SFHo and DD2. \textit{Right panel}: Corresponding posterior distributions of $R_{1.4}$ obtained under the minimal TOV mass, pQCD, gravitational-wave, and NICER constraints, but excluding the HESS measurement and the Gap object. For direct comparison, the prior distributions are overlaid, illustrating how these constraints reshape the allowed range of $R_{1.4}$.
}\label{fig:pQCD_BlWi-Rdist}
\end{figure}

In Fig.~\ref{fig:pQCD_BlWi-Rdist}, we present the distributions of the radius $R_{1.4}$ for a $1.4M_\odot$ NS. The left panel depicts the prior-induced spread of $R_{1.4}$ obtained from the hard prior on the equation of state, comparing two different hadronic phase models, SFHo and DD2. These prior distributions reflect only the physical requirements of stability and causality, without including any observational constraints. The right panel shows the resulting posterior distributions when the minimal TOV mass, pQCD, gravitational-wave, and NICER constraints are imposed, while HESS and Gap measurements are not taken into account. Overlaid prior distributions allow a direct visual comparison, highlighting how the applied constraints narrow and shape the range of allowed radii. In general, the peaks of the resulting distributions are in the range of 12.5 to 12.8 km.

\begin{figure}[h]
\centering
\includegraphics[width=0.5\textwidth]{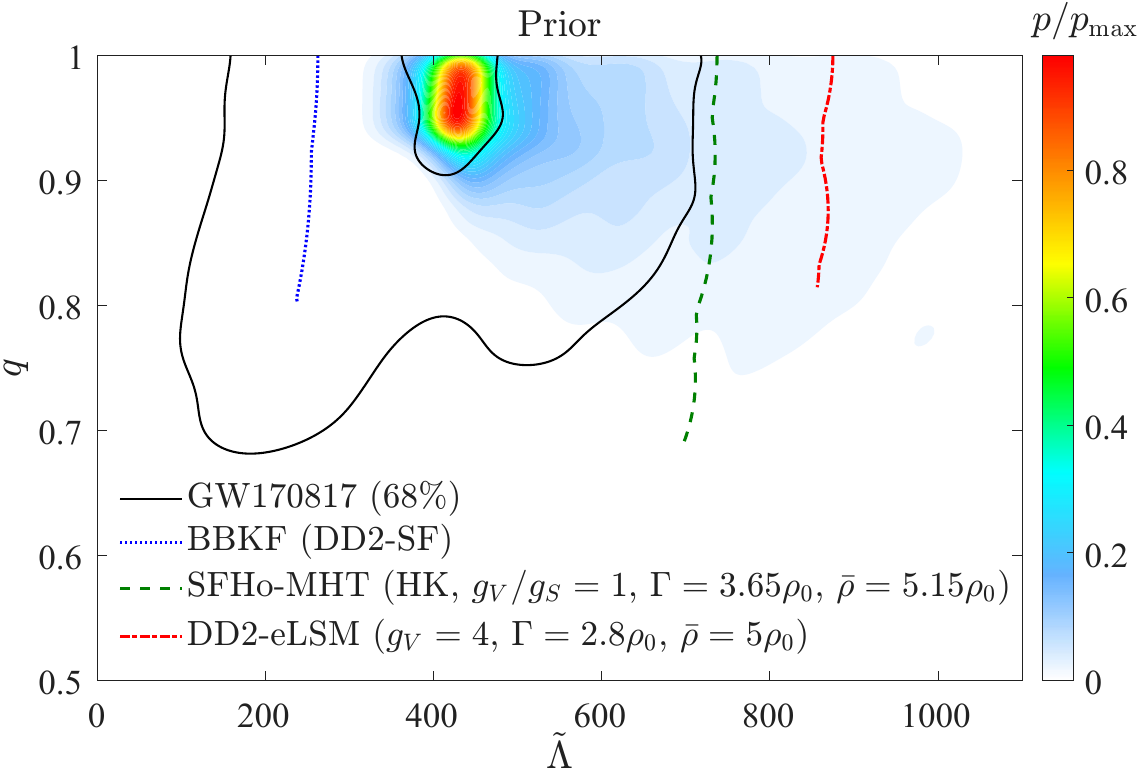}\hfill
\includegraphics[width=0.5\textwidth]{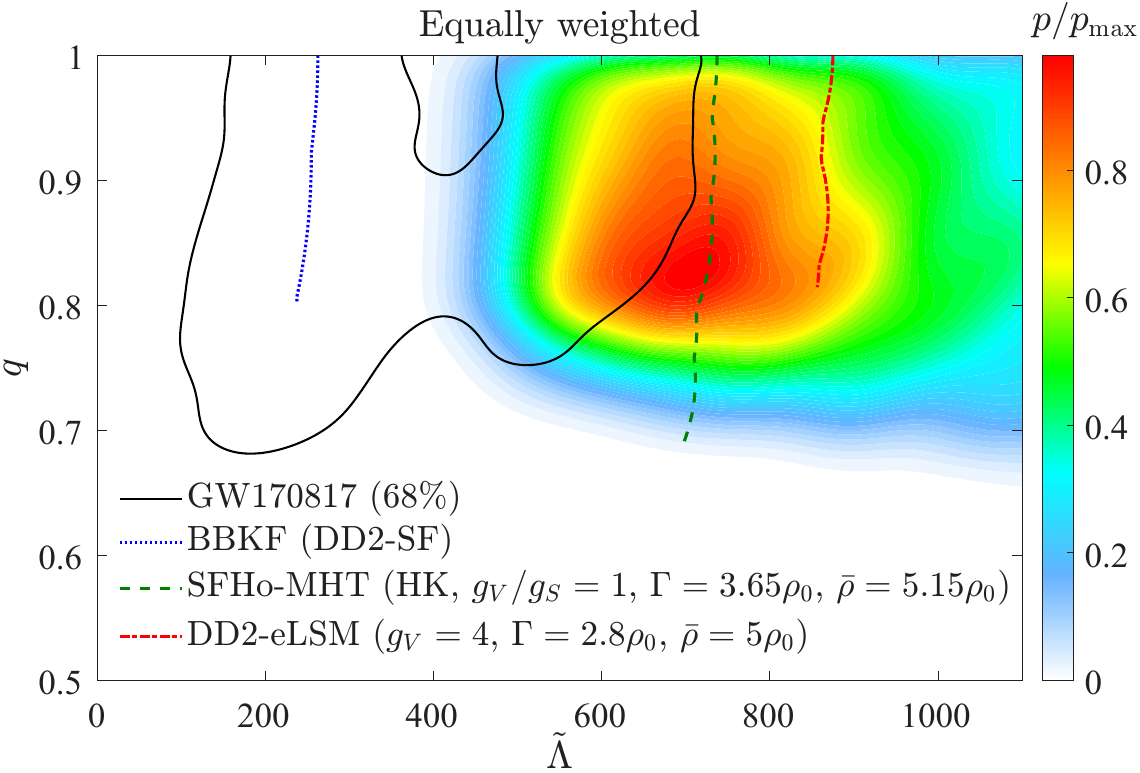}\hfill
\includegraphics[width=0.5\textwidth]{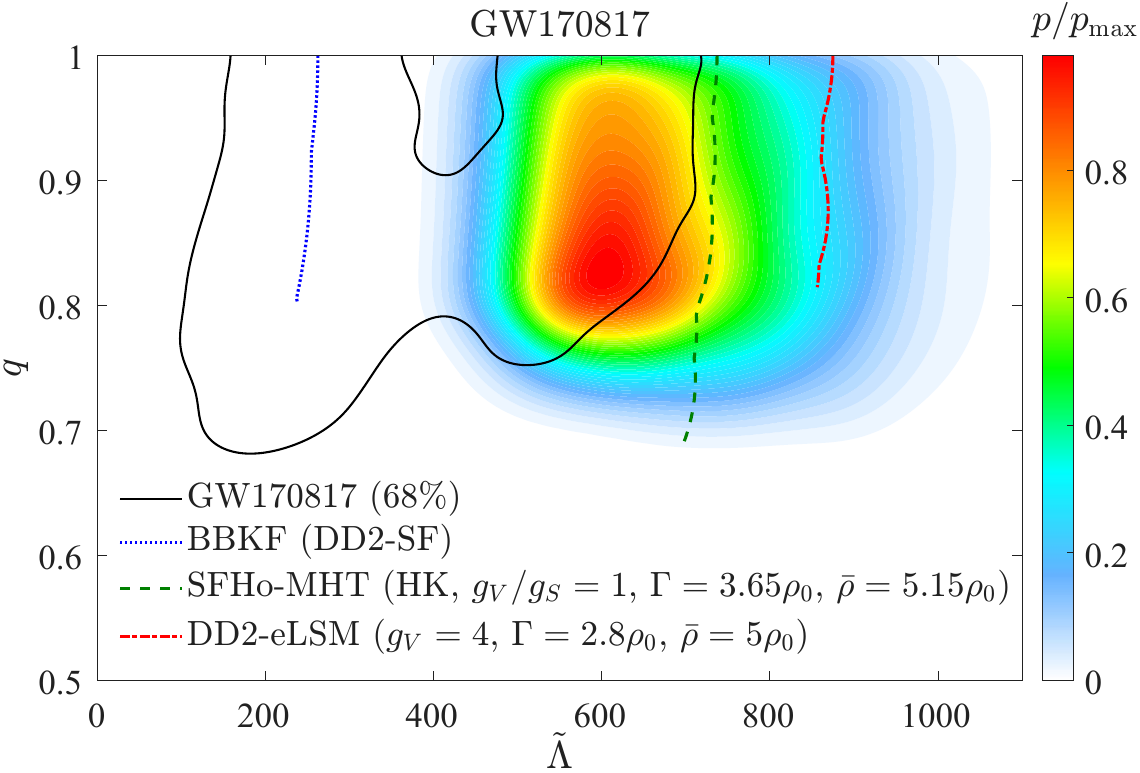}\hfill
\includegraphics[width=0.5\textwidth]{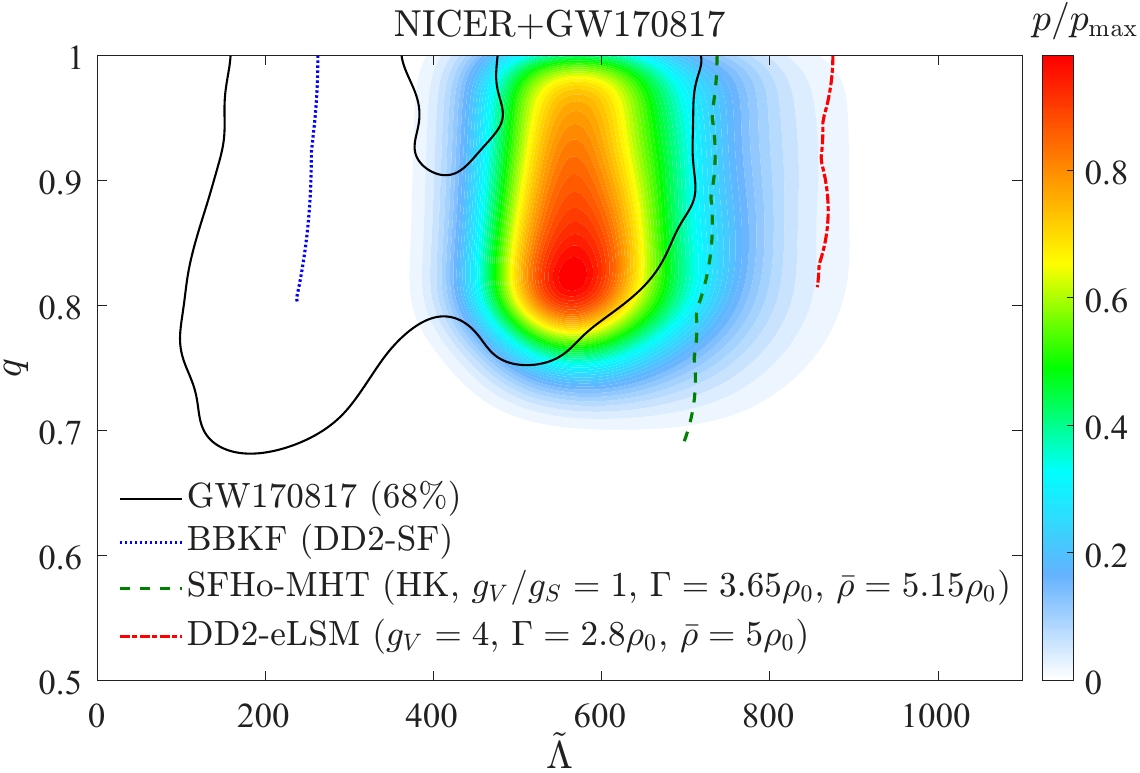}
\caption{The probability distribution in the $q$ (mass ratio) and $\tilde{\Lambda}$ (tidal deformability) plane. The upper left panel displays the density induced by the hard prior on the equation of state, obtained without likelihood weighting. In the upper right panel we only prescribe the minimal TOV mass and the pQCD constraint, in the bottom left figure we add the GW condition, while in the bottom right panel the NICER conditions are applied, too. The black line indicates the result of the LIGO–Virgo analysis of Ref.~\citep{LIGOScientific:2018hze}. In addition, the predictions corresponding to the three alternative equations of state considered in this work — BBKF (DD2–SF), SFHo–MHT, and DD2–eLSM — are overlaid as curves in all panels for comparison.}\label{fig:pQCD_BlWi-qLambda}
\end{figure}

Finally, to illustrate the impact of the tidal deformability on EOS, Fig.~\ref{fig:pQCD_BlWi-qLambda} presents the probability distribution in the plane of the mass ratio $q$ of the two merging NSs, and the mass-weighted linear combination of their tidal parameters $\tilde{\Lambda}$. The black line indicates the result of the LIGO–Virgo analysis at 68\% confidence level of Ref.~\citep{LIGOScientific:2018cki}. In the upper left panel we show the prior-induced density in the $(q,\tilde{\Lambda})$ plane, obtained by mapping the ensemble of equations of state allowed by the hard prior onto their corresponding $\tilde{\Lambda}$–$q$ relations, without applying any likelihood weighting. In the upper right panel, only the minimum-mass requirement and the pQCD constraint are imposed. In the bottom left panel, the GW constraint is added, while in the bottom right panel the NICER constraints are included as well. In addition to the posterior distributions obtained from our Bayesian analysis, we overlay the predictions of three alternative equations of state — BBKF (DD2–SF), SFHo–MHT, and DD2–eLSM — onto the $(q,\tilde{\Lambda})$ plane. For each of these models, the corresponding $\tilde{\Lambda}$–$q$ relations are computed and shown as curves, providing a direct comparison between the posterior support and the representative benchmark equations of state.

\begin{figure}[h]
\centering
\includegraphics[width=0.49\textwidth]{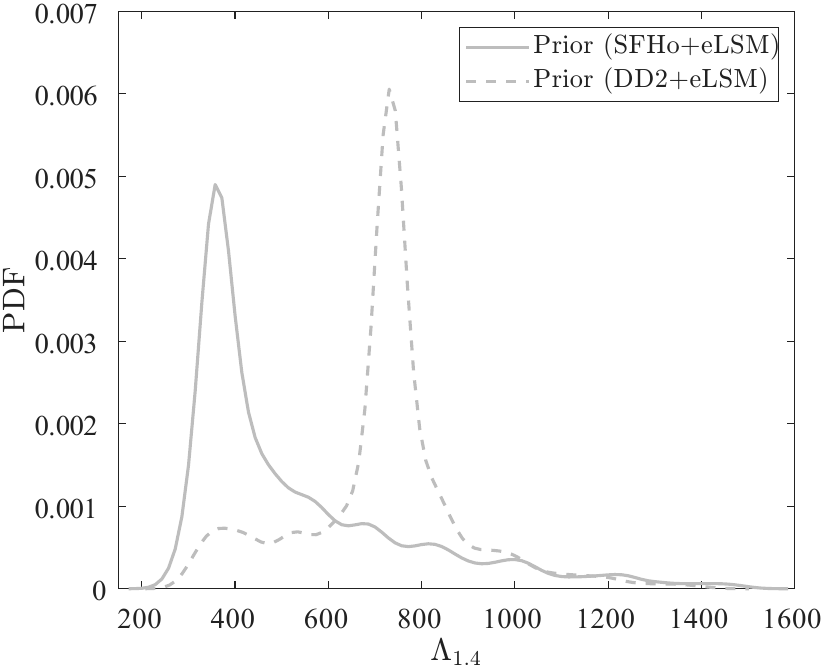}\hfill
\includegraphics[width=0.49\textwidth]{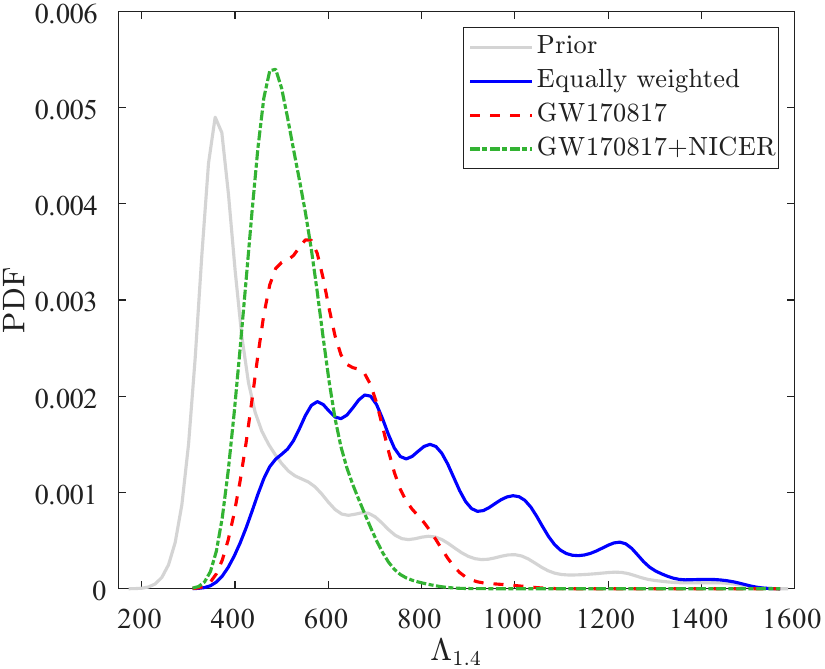}
\caption{\textit{Left panel}: Prior-induced distributions of the tidal deformability $\Lambda_{1.4}$ obtained from the hard prior on the equation of state, shown for two different choices of the hadronic phase: SFHo and DD2. \textit{Right panel}: Corresponding posterior distributions of $\Lambda_{1.4}$ inferred under the same set of constraints as in Fig.~\ref{fig:pQCD_BlWi-qLambda}. For reference, the prior distributions are overlaid in this panel to illustrate the impact of the observational constraints on the inferred tidal deformability.} \label{fig:pQCD_BlWi-Lambdadist}
\end{figure}

Fig.~\ref{fig:pQCD_BlWi-Lambdadist} shows the posterior distributions of the tidal deformability $\Lambda_{1.4}$ of a $1.4~M_\odot$ NS, obtained from our Bayesian analysis. In the left panel, we show the prior-induced distributions of $\Lambda_{1.4}$ resulting from the hard prior on the equation of state, for two different assumptions regarding the hadronic phase, namely the SFHo and DD2 models. These distributions reflect solely the physical constraints imposed by stability and causality, without incorporating any observational information. In the right panel, we present the corresponding posterior distributions obtained under the same set of observational constraints as those applied in Fig.~\ref{fig:pQCD_BlWi-qLambda}. For direct comparison, the prior distributions are overlaid, highlighting how the inclusion of the minimal TOV mass and the pQCD constraint, together with the gravitational-wave and NICER measurements, reshapes the allowed range of $\Lambda_{1.4}$.

Our results exhibit a preference for larger values of tidal deformability, in good agreement with the LIGO–Virgo analysis. In contrast, imposing the more restrictive constraint of Ref.~\citep{LIGOScientific:2018cki}, $70 < \Lambda_{1.4} < 580$, substantially reduces the number of EOSs compatible with the observational data by excluding a significant fraction of the $\Lambda_{1.4}$ distribution.

\section{Summary}

In this paper, we investigate the impact of individual NS observables on the allowed manifold of equations of state (EOSs) for strongly interacting matter. Constraints from low-energy nuclear physics were implemented using a chiral effective model that accurately describes low-density nuclear matter, whereas the asymptotically high-density regime was constrained by perturbative QCD. At intermediate chemical potentials, we employed a quark–meson model in which the vector coupling was treated as a free parameter, and the position and width of the transition between the low- and high-density regimes were varied to span a broad EOS manifold.

We considered a minimum value for the maximum NS mass of 
$M_\mathrm{TOV}^\mathrm{min}=2.22M_\odot$ \citep{Romani:2022jhd}, together with NICER mass–radius measurements for five NSs: PSR J0030+0451 \citep{Vinciguerra2024}, PSR J0740+6620 \citep{Salmi2024b}, PSR J0614–3329 \citep{Mauviard2025}, PSR J1231–1411 \citep{Salmi2024}, and PSR J0437–4715 \citep{Choudhury2024}. In addition, we incorporate constraints on the tidal deformability, adopting both the conservative bound $\tilde{\Lambda}<720$ \citep{LIGOScientific:2018hze} and the more restrictive range $70<\Lambda_{1.4}<580$ \citep{LIGOScientific:2018cki}.

Finally, we examined the impact of two hypothetical NSs: the central compact object in HESS J1731–347, with M=$0.77^{+0.20}_{-0.17}$~$M_\odot$ and R$_{1.4} = 10.4^{+0.86}_{-0.78}$~km, and the mass-gap object observed in GW190814.

We find that the constraint imposed on the minimum value of $M_{\mathrm{TOV}}$ has a particularly strong impact, retaining only about one third of the equations of state in the prior ensemble. In our analysis, this lower bound on $M_{\mathrm{TOV}}$ is based on a population-level estimate that includes PSR J0952–0607 (the black widow pulsar), rather than on the mass measurement of this source alone. As a consequence of this constraint, the equations of state considered in our analysis largely suppress NS radii below approximately 12 km. However, we also note that even the prior ensemble itself does not significantly populate radii below approximately 11 km, indicating that the prior already favours comparatively large radii. This leads to a tension between our model predictions and the 1$\sigma$ NICER contour for PSR J0614–3329. If the mass-gap object is indeed a NS, the EOS is constrained even more severely. In contrast, the properties inferred for the HESS J1731–347 object have little impact on the EOS. While NICER measurements provide valuable information, their current uncertainties imply that they do not yet significantly reshape the EOS probability distribution.

The tidal deformability emerges as another highly restrictive observable. 
Our model framework and the remaining observational data favor relatively large values of $\Lambda_{1.4}$. Although the conservative upper limit $\tilde{\Lambda}<720$ formally refers to the mass-weighted combination of the tidal deformabilities of the two merging NSs, it remains broadly compatible with the relatively large $\Lambda_{1.4}$ values favored by our model, whereas the more stringent constraint $70<\Lambda_{1.4}<580$ substantially reduces the allowed EOS manifold.

To describe these NS data, a very broad crossover transition between hadronic and quark matter is favoured in our model, with a transition density of approximately $4.8\rho_0$. In contrast, the hybrid equations of state constructed by Masuda et al.~\cite{Masuda:2012ed}, in which the NJL quark-matter EOS is matched to various hadronic models, typically predict a much narrower transition region with $\Gamma \simeq \rho_0$ and an onset density of $\overline{\rho} \simeq 3\rho_0$. In the present work, we adopt only the quark-matter EOS of Ref.~\cite{Masuda:2012ed} and match it to the SFHo hadronic EOS via a polynomial interpolation. Within this framework, we employ representative values of $\Gamma = 3.65\rho_0$ and $\overline{\rho} = 5.15\rho_0$, reflecting the typical scale of the crossover region in this specific construction. For comparison, the model including a first-order phase transition by Bastian et al.~\citep{Bastian:2020unt} exhibits a density jump at the transition from $1.5\rho_0$ to $3\rho_0$, and these values depend slightly on the electron fraction.

Understanding why the present class of models prefers higher values of $\Lambda_{1.4}$ is an important open question that we plan to address in a forthcoming study. Another natural extension of this work is the inclusion of strangeness degrees of freedom in the hadronic phase, which may further modify the EOS at high densities.

\addcontentsline{toc}{section}{Acknowledgements}\bmhead{Acknowledgements}

This work was supported by the Hungarian OTKA fund K138277.
G.K. acknowledges support from the KKP-2024 Research Excellence Programme of MATE, Hungary. 
J.T. is supported by the Alexander von Humboldt Foundation under the project no. 1240213 - HFST-P.

\bibliography{eLSM_bayes}

\end{document}